\documentclass[aip,graphicx]{revtex4-1}
\usepackage{graphicx,siunitx,tabularx} 

\draft 
\renewcommand{\selectlanguage}[1]{}
\begin{document}


\title{Seedless magnetic field generation via ultraintense laser irradiation of an aluminum tetrafoil} 



\author{Jessa Jayne C. Miranda}
\email[]{jmiranda@nip.upd.edu.ph}
\affiliation{National Institute of Physics, University of the Philippines Diliman, Quezon City 1101, Philippines}

\author{Myles Allen H. Zosa}
\affiliation{National Institute of Physics, University of the Philippines Diliman, Quezon City 1101, Philippines}


\date{\today}

\begin{abstract}
Magnetic field generation and amplification techniques are popular due to their applications in several fields, including laboratory astrophysics and inertial confinement fusion. Microtube Implosion (MTI) is a scheme that amplifies a seed magnetic field by several orders of magnitude through ultraintense laser irradiation of solid targets. In this study, we conducted 2D EPOCH simulations to investigate the ability of a tetrafoil target to generate a magnetic field in the absence of a seed magnetic field. The results show that the tetrafoil target generates magnetic field intensities comparable to a hollow cylindrical aluminum target undergoing MTI with a \SI{10}{\kilo\tesla} seed. Furthermore, the configuration of the tetrafoil can be used to control the direction of the magnetic fields produced. The combined tetrafoil and microtube target sustained higher magnetic field and flux compared to both the microtube and tetrafoil targets, without the need for a seed magnetic field.
\end{abstract}

\pacs{}

\maketitle 

\section{Introduction}
Researchers have become increasingly interested in generating and amplifying strong magnetic fields, driven by the potential applications of intense magnetic fields in various research areas, including laboratory astrophysics \cite{Intro-Astro-Remington2006, Intro-Astro-Bulanov2015, Intro-Astro-Revet2021} and inertial confinement fusion (ICF) \cite{Intro-Inertial-Perkins2017, Intro-Inertial-Sio2023, Intro-Inertial-Yager2022}. Magnetic fields play an important role in laboratory astrophysics specifically for magnetic reconnection, magnetic turbulence, and collisionless shock experiments \cite{Intro-Astro-Takabe2021}. A well-known theory on the origin of the strong magnetic fields in space is that of the turbulent dynamo, wherein a seed field is amplified by the hydrodynamic mechanism in plasmas \cite{Intro-Astro-Gregori2015, Intro-Astro-Laishram2024}. Several studies have been done to determine how the seed field is generated in the first place, with the Biermann Battery Effect as one of the major candidates  \cite{Intro-Astro-Ohira2021, Intro-Astro-Shukla2020}. On the other hand, in ICF studies, magnetic fields have been used to ease the ignition requirements by regulating the thermal conduction losses and confining charged fusion products \cite{Intro-Inertial-Wurden2016, Intro-Inertial-Yager2022}. In 2023, Sio et al. reported that applying magnetic fields of \SI{12}{\tesla} and \SI{26}{\tesla} increased the deuterium-deuterium neutron yield by 150\% and 190\%, respectively, in a cryogenic deuterium-tritium ICF experiment\cite{Intro-Inertial-Sio2023}. 

Various studies developed different schemes to generate and amplify magnetic fields with some using relativistic laser pulses (laser intensities $>$ \SI{e18}{\watt\per\square\centi\meter}) \cite{Intro-MagFieldGeneration-Jiang2021, Intro-MagFieldGeneration-Longman2021, Intro-MagFieldGeneration-Peterson2021}. These techniques primarily employ the inverse Faraday effect (IFE) to generate an axial magnetic field by transferring angular momentum from a laser pulse to the plasma. The laser-plasma interactions causes the compression of the magnetic flux, thus amplifying the magnetic field to gigagauss level for less than a picosecond \cite{Intro-MagFieldGeneration-Jiang2021}. 

In 2020, Murakami et al. proposed a scheme called Microtube Implosion (MTI), that can generate megatesla (MT)-order magnetic fields by placing a microtube target under a constant seed magnetic field and irradiating it with ultraintense laser pulses. In MTI, the laser pulses heat up the electrons within the target, causing them to quickly form an electron sheath on the target-cavity interface. The electron sheath pulls the nearby ions, triggering an implosion. As the electrons and ions implode, the seed magnetic field deflects their trajectories, creating a Larmor hole at the center. The current due to the imploding plasma results in an amplified magnetic field much stronger than the seed, reaching MT-order  \cite{Intro-MTI-Murakami2020}. Gu and Murakami extended this study by varying the seed magnetic fields. They determined that for values $>$ \SI{250}{\tesla}, the maximum magnitude and the time evolution of the resulting magnetic fields are similar \cite{Intro-MTI-Gu2021}. Zosa et al. proposed a target structure that can generate \SI{100}{\kilo\tesla} magnetic field without needing a seed \cite{Intro-MTI-Zosa2022}. They observed that using both a paisley target and a double paisley target result in the generation of kT-level magnetic fields, with the latter extending the magnetic field lifetime to ps-scale \cite{Intro-MTI-Zosa2022}. Thus, variations in the target structure may be useful for MTI by removing the need for a seed magnetic field. 

In this study, we investigate the magnetic field generation capabilities of a novel target configuration via 2D EPOCH simulations. To assess how this compares to the magnetic field generated through MTI, we also simulated the MTI of an aluminum microtube as a baseline. 

\section{2D EPOCH Simulation}
Particle-in-cell (PIC) simulations are used to gain insights on processes that occur in ultra-short time scales \cite{Simulation-Gallman2012}. EPOCH is a PIC implementation that is commonly used to investigate magnetic field generation. In this study, we used 2D EPOCH simulations to model the laser-plasma interaction \cite{Simulation-Arber2015}. A previous study has shown that for MTI, the 2D and 3D simulations agree well with each other\cite{Intro-MTI-Shokov2021}. Thus, the 2D results can be a good approximation to a 3D system.  Additionally, EPOCH simulations have been shown to be consistent with experimental findings\cite{Simulation-Law2020}. 

\begin{figure}
\includegraphics{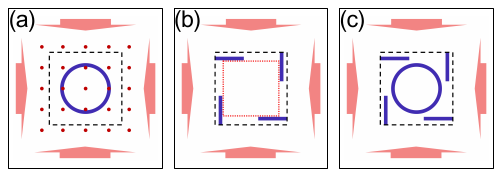}
\caption{Schematic of the simulation setups used in this study, showing the lasers (red arrows), seed magnetic field (red dots), and the aluminum target (blue shapes).}
\label{Fig:Schematic}
\end{figure}

The schematics of the 2D EPOCH simulation setups used in this study are shown in Fig.~\ref{Fig:Schematic}. 
For the baseline simulation (Fig.~\ref{Fig:Schematic}a), four p-polarized lasers (red arrows) each with wavelength of $\lambda_L$ = \SI{1}{\micro\meter}, Full-width-half-maximum (FWHM) of $\tau_L$ = \SI{100}{\femto\second}, and intensity of $I_L$ = \SI{e21}{\watt\per\square\centi\meter}, originated from the four sides of the \SI{40}{\micro\meter} $\times$ \SI{40}{\micro\meter} simulation box. The simulation area was divided into 4000 $\times$ 4000 cells. Each cell occupied by the target contained 100 aluminum ions and 200 electrons. All targets have an initial density of \SI{e22}{\per\cubic\centi\meter}. The inner radius of the \SI{0.50}{\micro\meter} thick microtube (hollow) target was \SI{3}{\micro\meter}. The target was placed under a \SI{10}{\kilo\tesla} seed magnetic field along the +z-direction (red dots). The total simulation time was \SI{1}{\pico\second} with the ion density and magnetic field data recorded every \SI{10}{\femto\second}.

We replaced the hollow target with a tetrafoil (TF) composed of four \SI{0.5}{\micro\meter} thick aluminum foils that are \SI{4}{\micro\meter} long and positioned as shown in Fig.~\ref{Fig:Schematic}b. The red dotted square represents the region enclosed by the foil. The mirrored tetrafoil (MTF) target uses the same parameters as the TF target but are placed in a configuration mirroring Fig.~\ref{Fig:Schematic}b. We also tested the combination of the microtube and TF targets (Fig.~\ref{Fig:Schematic}c) to determine if the TF target is a viable alternative to a seed field in MTI. 

The ion density and magnetic field data presented in this paper are confined to the region enclosed by the black dashed squares in Fig.~\ref{Fig:Schematic}. Due to the high laser intensities, the plasma is collisionless \cite{Simulation-Miloshevsky2022}. To lessen the computational cost, the targets were pre-ionized, consisting of fully ionized aluminum. 

\section{Results and Discussion}

\begin{figure}
\includegraphics{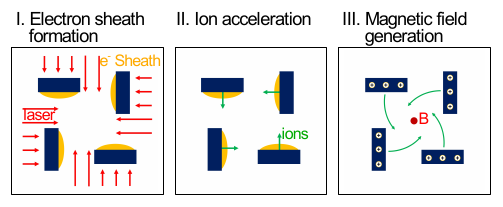}
\caption{Seedless magnetic field generation mechanism using a tetrafoil (TF) target.}
\label{Fig:TF-Mech}
\end{figure}

A highly localized intense magnetic field is produced via MTI of a hollow aluminum target. However, this relies on a seed magnetic field.  Our goal is to investigate whether it is possible to produce a localized magnetic field without a seed magnetic field. Figure~\ref{Fig:TF-Mech} shows the schematic of the seedless magnetic field generation using a tetrafoil (TF) target. It starts with four p-polarized lasers illuminating the TF target shown in Fig.~\ref{Fig:Schematic}b. The fast electrons from the front (laser-facing) surfaces quickly propagate through the foils, accumulating at the rear surfaces and forming electron sheaths. These sheaths generate electric fields that accelerate the ions normal to the rear surfaces. As the sheath dissipates, the lasers and the positively charged foils deflect the slower ions, producing a counterclockwise current, resulting in the formation of a magnetic field in the +z-direction which is amplified when the subsequent ions implode. 

\begin{figure}
\includegraphics[width=0.7\linewidth]{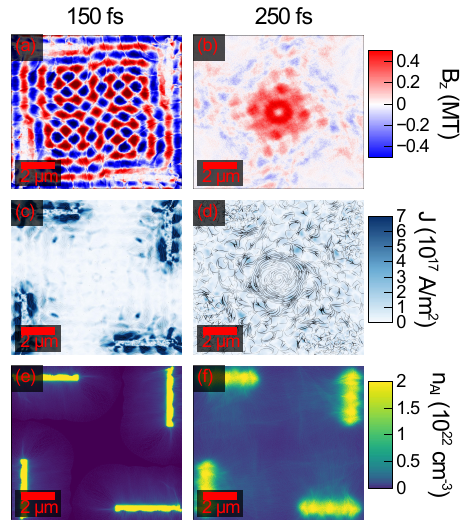}
\caption{Magnetic field profile (a,b), current density (c,d), and ion density (e,f) of the TF target at 150 and \SI{250}{\femto\second}.}
\label{Fig:TF-sim}
\end{figure}

Figure~\ref{Fig:TF-sim} shows the magnetic field profile (a,b), current density (c,d), and ion density (e,f) of the TF target at 150 and \SI{250}{\femto\second}. When the lasers initially interact with the target, the magnetic field profile follows that of the lasers (Fig.~\ref{Fig:TF-sim}a). As the lasers transfer their energies to the foils, intense surface currents are produced on their surfaces (Fig.~\ref{Fig:TF-sim}c). The electron sheaths formed at the rear surfaces accelerate the ions toward the center as the laser and the nearby foils curve the ion trajectory (Fig.~\ref{Fig:TF-sim}e). As more ions implode (Fig.~\ref{Fig:TF-sim}f), the intensity of the spin currents increase (Fig.~\ref{Fig:TF-sim}d), producing a strong  localized field along the +z-direction as seen in Fig.~\ref{Fig:TF-sim}b.

\begin{figure}
\includegraphics{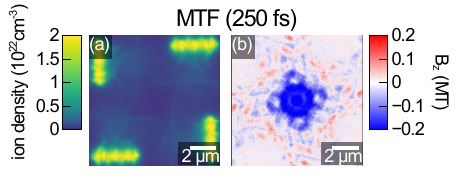}
\caption{Ion density (a) and magnetic field profile (b) of the MTF target at \SI{250}{\femto\second}.}
\label{Fig:MTF}
\end{figure}

The aluminum ion density of the MTF target and its corresponding magnetic field profile at \SI{250}{\femto\second} are shown in Fig.~\ref{Fig:MTF}. We can see that the MTF target also generated a strong localized magnetic field albeit in the opposite direction as the TF target. This implies that the magnetic field direction can be controlled by the appropriate positioning of the tetrafoil. 

\begin{figure}
\includegraphics{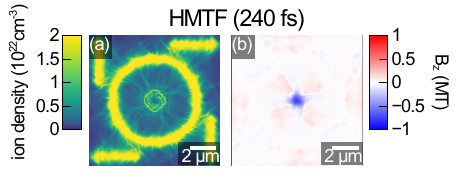}
\caption{Ion density (a) and magnetic field profile (b) of the HMTF target at \SI{250}{\femto\second}.}
\label{Fig:HMTF}
\end{figure}

To check if this can be extended to a combination of a tetrafoil and hollow target (HTF, HMTF), we also performed simulations using the HTF and HMTF targets shown in Fig.~\ref{Fig:Schematic}c. The aluminum ion density and magnetic field profile of the HMTF target at \SI{250}{\femto\second} are shown in Fig.~\ref{Fig:HMTF}. The HMTF target generated a strong localized magnetic field with a relatively smaller area compared to that of the MTF target. This means that the MTF target is preferable if we wish to employ the magnetic fields for applications in need of larger area coverage. 

\begin{figure}
\includegraphics[width=0.75\linewidth]{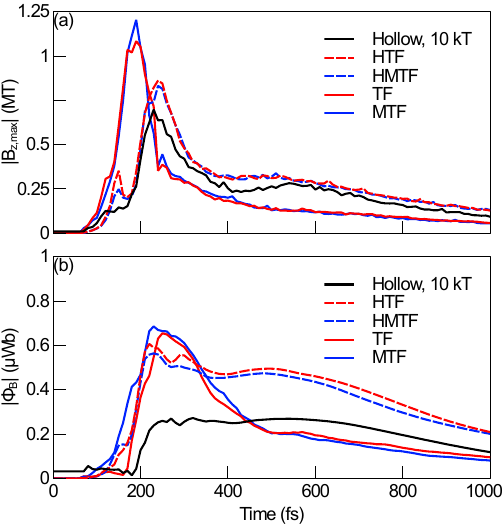}
\caption{Time evolution of the maximum magnetic field and magnetic flux at the center of the tetrafoil target (MTF and TF) and combined hollow and tetrafoil targets (HMTF, HTF).}
\label{Fig:TF-MTF-BZprof}
\end{figure}

To examine if adding a tetrafoil structure to the microtube target can enhance the properties of the generated magnetic field, we investigate the time evolution of the maximum magnetic field and flux at the center of the different targets as shown in Fig.~\ref{Fig:TF-MTF-BZprof}. The results clearly show that the structured targets generate magnetic fields that are stronger than that of the seeded hollow target, reaching MT-order for approximately \SI{30}{\femto\second} for the TF and MTF targets. However, due to their open structure, the laser traverse the central area of the TF and MTF targets, which inflate their maximum magnetic fields in Fig.~\ref{Fig:TF-MTF-BZprof}. After \SI{200}{\femto\second}, the maximum magnetic fields of both the TF and MTF targets quickly drop to below that of the seeded hollow target but still maintain 100kT-level magnetic fields. On the other hand, the maximum magnetic field and magnetic flux inside the target cavity of the HTF and HMTF targets closely resemble that of the seeded hollow target suggesting that adding the tetrafoil eliminated the need for a seed magnetic field. The magnetic flux of the TF and MTF targets are significantly higher than the seeded hollow target until around \SI{400}{\femto\second}, after which their magnetic flux become smaller than the seeded hollow target. Their magnetic flux is smaller until the end of the simulation. These results also show that the tetrafoil target can be a viable substitute to a seed magnetic field. This can be confirmed by looking at the magnetic flux of the HTF and HTMF targets that are more than 2x that of the TF and MTF targets as shown in Fig.~\ref{Fig:TF-MTF-BZprof}b. 

The summary of the $B_z$ results for all the target structures used in this study are shown in Table~\ref{Tab:BzSummarytable}. The maximum magnetic field inside the cavity or foil-enclosed region ($|B_{z,max}|$) is recorded with the time ($t_{max}$) at which it was observed. The direction of the localized magnetic field ($B_z$ direction) and time interval by which it sustains a magnetic field greater than \SI{0.5}{\mega\tesla} (Lifetime) are also noted. As we can see, the strongest maximum magnetic field magnitude was generated by the MTF target. However, as evidenced by Fig.~\ref{Fig:TF-MTF-BZprof}a and the Lifetime, this only lasts until around \SI{200}{\femto\second}. After that, the Hollow, HTF, and HMTF targets are seen to consistently outperform the TF and MTF targets in terms of maximum magnetic field. Furthermore, the direction of the localized magnetic field can be controlled by the appropriate positioning of the foils in the target as listed in the $B_z$ direction. 

\begin{table}
\caption{Summary of $B_z$ results for various target structures}
\label{Tab:BzSummarytable} 
\setlength{\tabcolsep}{5pt} 
\begin{tabular}{l|c c c c c}
\hline\hline
& \multicolumn{5}{c}{\textbf{Target Structure}}\\
& Hollow & TF & MTF & HTF & HMTF\\
\hline
$|B_{z,max}|$ $(MT)$ & 0.70 & 1.08 & 1.20 & 0.86 & 0.83\\
$t_{max}$ $(fs)$ & 230 & 190 & 190 & 240 & 240 \\
$B_z$ direction & + & + & - & + & -\\
Lifetime $(fs)$ & 60 & 80 & 80 & 90 & 80 \\
\hline\hline
\end{tabular}
\end{table}

\section{Conclusion}
We evaluated the performance of different target structures in generating magnetic fields via 2D EPOCH simulations. The results show that a tetrafoil target can produce MT-order magnetic fields without a seed magnetic field. The tetrafoil is shown to be a viable alternative to a seed magnetic field for MTI. Appropriate positioning of the foils can also be used to control the direction of the generated field. Further investigation on the effects of size and configuration is still needed to determine if the lifetime can be significantly improved. It would also be worthwhile to determine if these results can be replicated using a two- or single-laser system.

\begin{acknowledgments}
This research was supported by grants from the UP-OVPAA Balik PhD Program (Project No.OVPAA-BPhD-2022-07) and DOST-PCIEERD (Project No.11324). The simulations were performed using EPOCH (developed under UK EPSRC Grants EP/G054950/1, EP/G056803/1, EP/G055165/1, and EP/M022463/1).
\end{acknowledgments}

\bibliography{references.bib}

\end{document}